# Surface-mediated frequency aging beyond quality-factor saturation in an AlScN-on-silicon resonator

S M Enamul Hoque Yousuf, Mahmudul Hasan, Shruti Mishra, Sourav Mukherjee, Jinghan Gao, Roozbeh Tabrizian*

*Electrical Engineering and Computer Science, University of Michigan, Ann Arbor, MI 48109, USA*

## Abstract

Vacuum package integrity in micro- and nanoelectromechanical resonators is commonly assessed through the quality factor *Q*, although *Q* probes residual-gas damping rather than the surface-state evolution that can govern frequency aging. Here, we disentangle the pressure responses of *Q* and the resonance frequency $f_0$ in a 64.21 MHz $Al_{0.7}Sc_{0.3}N$-on-silicon cross-sectional Lamé-mode resonator between 0.01 to 760 Torr. Measurements at 25 °C and at the 68.8 °C frequency turnover, where first-order thermal sensitivity is suppressed, reveal widely separated equilibration timescales. Following each pressure step, *Q* reaches a reversible, history-independent steady value on the pressure-control timescale and approaches a fitted pressure-independent ceiling of $8.5\times10^4$ below approximately 1 Torr. By contrast, $f_0$ responds measurably down to $10^{-5}$ Torr and remains history-dependent, relaxing for hours at fixed pressure. The transients follow stretched-exponential kinetics, consistent with a broad distribution of surface relaxation rates, and individual pressure steps produce fractional frequency shifts as large as 1.9 ppm. Ten-hour phase-locked measurements show a common short-term time-deviation floor near $1\times10^{-10}$ s from 0.01 to 100 Torr, whereas residual deterministic relaxation dominates at long averaging times; at 760 Torr, gas damping degrades short-term tracking through the reduced *Q*. These results establish gas-damping equilibrium and frequency equilibrium as distinct states. Quality-factor saturation alone is therefore insufficient to qualify vacuum packaging for precision mechanical frequency references; package specifications must also constrain surface-mediated frequency aging.

***Keywords***: AlScN-on-silicon resonator; vacuum packaging; surface adsorption; frequency aging; gas damping; quality-factor saturation; frequency stability.

*Corresponding Author: rtabrizi@umich.edu

## 1. Introduction

Resonant micro- and nanoelectromechanical systems (M/NEMS) serve as frequency-selective building blocks in oscillators, filters, and sensors.[1–3] They are particularly critical in positioning, navigation, and timing systems, where local clocks must provide accurate holdover when satellite-based references are interrupted or denied. Communications, radar, and distributed electronic sensing impose similar requirements through their dependence on precise synchronization.[4] Chip-scale atomic clocks have brought precision timekeeping to battery-powered platforms,[5] but mechanical resonators remain attractive for applications subject to stringent size, weight, power, and cost constraints.

Recent advances are pushing micromechanical frequency references towards stability levels historically associated with atomic standards.[6,7] At short averaging times, the high quality factor ($Q$) of M/NEMS resonators supports low phase noise, while doping-based temperature compensation and ovenized operation suppress thermally induced frequency drift. [8–10] As these sources of instability are progressively reduced, however, extended holdover becomes constrained by aging-the slow, deterministic evolution of the resonance frequency $f_0$ that accumulates into timing error over hours, days, and longer intervals. Identifying and controlling the physical mechanisms responsible for aging is therefore essential for translating the excellent short-term stability of micromechanical clocks into holdover performance approaching that of atomic standards.

A mechanical resonance is principally characterized by its resonance frequency $f_0$, which determines the oscillation rate, and its quality factor $Q$, which quantifies energy retention. The quality factor can be expressed as $Q=2\pi E_{\text{stored}}/\Delta E_{\text{cycle}}$, and independent dissipation mechanisms contribute additively through $Q^{-1}=\Sigma_i Q_i^{-1}$. Systematic suppression of these loss mechanisms has enabled micro- and nanomechanical resonators with increasingly high $Q$, in some cases approaching material-imposed fundamental limits.[11–13] In shear bulk acoustic wave modes, such as the Lamé mode, the volume-preserving (isochoric) deformation suppresses thermoelastic dissipation,[11,14] while phononic bandgap structures suppress anchor loss by inhibiting elastic-wave radiation into the substrate.[15] Once these major loss channels are sufficiently suppressed, anharmonic interaction of the acoustic mode with background phonons imposes a material-dependent $f_0Q$ ceiling known as the Akhiezer limit.[11,16]

Translating this low-loss mechanical response into a practical frequency reference additionally requires efficient electromechanical transduction and control of thermal sensitivity. Integrated piezoelectric films provide large electromechanical coupling[1,17,18] and enable all-electrical excitation and read-out of tens-of-megahertz Lamé modes engineered at the cross-section of silicon waveguides.[8] The use of electronic effects through heavy doping of silicon enables tailoring the temperature dependence of the frequency defining elastic constants, producing a turnover temperature at which the first-order temperature coefficient of frequency ($TCF_1$) vanishes.[9,10] Together, these advances provide resonators with low pressure-independent dissipation and a first-order temperature-compensated operating point, creating a controlled platform for investigating mechanisms that govern long-term frequency evolution.

As bulk, thermal, and support-related limitations are reduced, surface and package-mediated processes become increasingly important, particularly as miniaturization increases the surface-to-volume ratio.[19] Long-term oscillator aging has historically been associated with mass transfer at surfaces and interfaces within sealed packages.[20] Wafer-level vacuum encapsulation is commonly qualified by monitoring $Q$, with long-term stability of $Q$ interpreted as evidence that the package pressure remains stable.[21,22] This criterion, however, does not necessarily establish that the resonance frequency has equilibrated. Residual gas primarily influences $Q$ through dissipative loading, whereas $f_0$ responds to changes in both the effective modal mass and the effective stiffness. Adsorption, desorption, and reorganization of surface species can therefore shift $f_0$ even after pressure-dependent gas damping has reached steady state and $Q$ has saturated. Previous studies have examined these processes largely in isolation: pressure-dependent dissipation,[21–25] adsorption-desorption noise,[26,27] finite-site adsorption processes,[28] atomically thin membranes,[29] and excess frequency fluctuations in nanomechanical resonators.[30] Surface adsorption and air damping have also been investigated together in β-$Ga_2O_3$ nanomechanical resonators.[31] Nevertheless, a unified time-resolved measurement that directly compares dissipation equilibrium, frequency relaxation, and closed-loop frequency stability from high vacuum to atmospheric pressure, while suppressing first-order thermal sensitivity, remains lacking. Consequently, the vacuum level inferred from $Q$ may differ substantially from that required to achieve long-term frequency stability.

Here, we investigate the pressure-dependent dissipation and frequency evolution of a 64.21 MHz $Al_{0.7}Sc_{0.3}N$-on-silicon (AlScN-on-Si) cross-sectional Lamé-mode resonator. Measurements are performed at 25 °C and at the 68.8 °C frequency turnover, where $df_0/dT$ vanishes to first order. We vary the chamber pressure over nearly five decades, from 0.01 to 760 Torr, and combine resonance-spectrum measurements, repeated pressure cycling, a high-vacuum pump-isolation transient, hour-scale frequency-relaxation measurements, and ten-hour phase-locked frequency records. The quality factor reaches a reversible, history-independent steady state and approaches its pressure-independent ceiling below approximately 1 Torr. In contrast, $f_0$ responds measurably down to approximately $10^{-5}$ Torr, remains dependent on pressure history, and continues to evolve for hours after the chamber pressure has stabilized. The frequency transients follow stretched-exponential kinetics with stretching exponents below unity, indicating a broad distribution of surface relaxation rates, and individual pressure steps produce fractional frequency shifts as large as 1.9 ppm. Closed-loop time-deviation measurements show a common short-term floor near $1\times10^{-10}$ s from 0.01 to 100 Torr, whereas residual deterministic relaxation governs the response at longer averaging times. At atmospheric pressure, gas damping reduces $Q$ and degrades short-term frequency tracking. These observations establish gas-damping equilibrium and frequency equilibrium as distinct states. Quality-factor saturation alone is therefore insufficient to qualify vacuum packaging for precision mechanical frequency references; package design and qualification must also constrain surface-mediated frequency aging over the intended operating and holdover intervals.

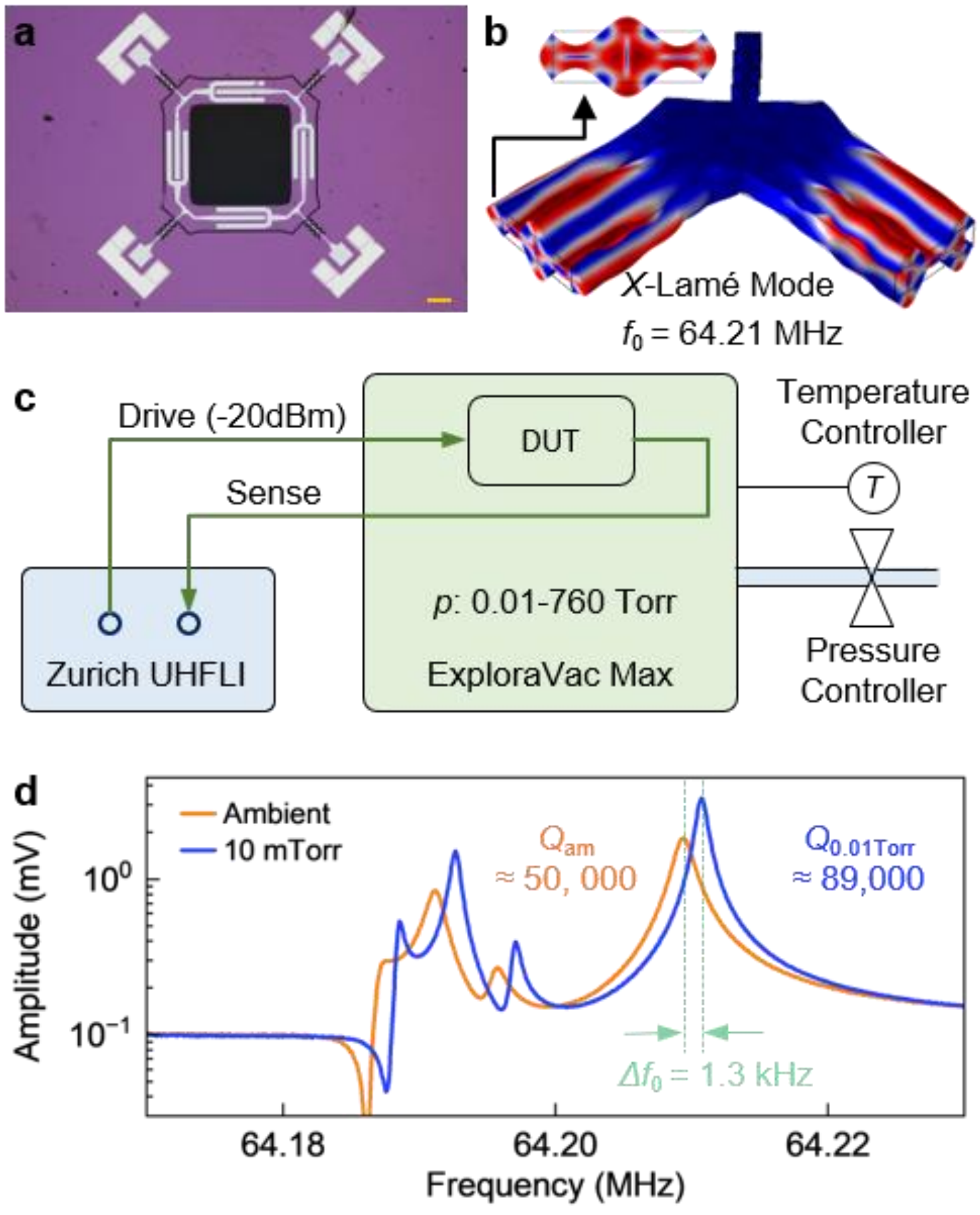

**Figure 1. AlScN-on-Si Lamé-mode resonator and pressure-dependent response.** (a) Optical micrograph of the resonator. The phosphorus-doped silicon body is suspended by phononic crystal tethers inside a 10 µm sidewall gap, and the top molybdenum (Mo) electrode is patterned on the AlScN film to provide all-electrical actuation and readout. Scale bar: 150 µm (b) Simulated mode shape of the third order out-of-plane cross-sectional Lamé mode at 64.21 MHz. (c) Schematic of the measurement system showing the device under test (DUT) inside an ExploraVac Max chamber with closed-loop pressure and temperature control and is driven and read out by a Zurich UHFLI lock-in amplifier. (d) Transmission spectra measured at 25 °C and -20 dBm drive, at ambient pressure (orange) and 10 mTorr (blue). Reducing the pressure increases $f_0$ by approximately 1.3 kHz. Neighboring peaks near 64.19 MHz correspond to adjacent arms of the same resonator; the strongest mode at 64.21 MHz is tracked throughout this work.

## 2. Results

### Design and Characterization of the X- Lamé Mode Resonator

**Figure 1**a shows an optical image of the degenerately phosphorus-doped single-crystal silicon cross-sectional Lamé-mode (X-Lamé) resonator. The resonator body is 600 µm long and has a 150 µm × 50 µm cross section. It is separated from the surroundings by a 10 µm gap, supported at its corners by phononic-crystal tethers, and released from the backside. Actuation and read-out are all-electrical, though a piezoelectric stack patterned on the top surface: a 600 nm c-axis-textured

$Al_{0.7}Sc_{0.3}N$ film and a 50 nm molybdenum layer defining separate drive and sense electrodes, with the low-resistivity silicon body serving as the bottom electrode. The transducer excites the third-order out-of-plane X-Lamé mode at $f_0$ = 64.21 MHz, whose simulated displacement field appears in **Figure 1**b. Alternating displacement lobes extend across the resonator body, while the motion is predominantly normal to the top surface, so the sidewall gap is not periodically compressed and squeeze-film damping is suppressed. Residual gas instead interacts primarily with the exposed surfaces of the resonator body, whose 150 µm width defines the characteristic length $L$ used in the Knudsen number analysis. The volume-preserving shear of the Lamé mode suppresses thermoelastic dissipation, while attachment at the quasi-nodal corners reduces coupling to the supports and the periodic phononic cells reflect radiated elastic energy back into the resonator body, thereby minimizing anchor loss. Phosphorus doping establishes a $f_0(T)$ turnover at 68.8 °C. Together, these design features suppress major loss channels and establish a first-order temperature-compensated operating point, providing a controlled platform for separating pressure-dependent dissipation from surface-mediated frequency evolution.

**Figure 1**c shows the measurement platform: the wire-bonded die sits inside an ExploraVac Max vacuum chamber with independent closed-loop control of pressure (0.01-760 Torr) and temperature, while a Zurich Instruments UHFLI lock-in amplifier drives the transducer and records the transmission spectrum at each pressure and temperature set point. Venting the chamber from 10 mTorr to ambient pressure reduces the resonance amplitude by approximately half and decreases $f_0$ by roughly 1.3 kHz, as shown in **Figure 1**d. At a fixed electrical drive, the reduction in resonance amplitude indicates increased dissipative loading by the surrounding gas. The frequency shift can arise from changes in the effective modal mass or stiffness; the observed decrease in $f_0$ is consistent with increased adsorbate mass loading at higher pressure.

**Pressure Responses of Dissipation and Resonance Frequency**

The pressure-dependent resonance response is governed by changes in effective mass, effective stiffness, and energy dissipation. For a resonator body with effective mass $m_{\mathrm{eff}}$ and effective spring constant $k_{\mathrm{eff}}$, the natural frequency is expressed as $\omega_0=2\pi f_0=(k_{\mathrm{eff}}/m_{\mathrm{eff}})^{1/2}$. A small perturbation therefore produces a fractional frequency shift given by $\Delta f_0/f_0 \approx (1/2) \times (\Delta k_{\mathrm{eff}}/k_{\mathrm{eff}} - \Delta m_{\mathrm{eff}}/m_{\mathrm{eff}})$. Adsorbed mass lowers $f_0$, whereas adsorbate-induced surface stress can shift $k_{\mathrm{eff}}$. The frequency shift alone does not uniquely distinguish these contributions, although a decrease in $f_0$ is consistent with mass loading when the stiffness contribution is small. Energy loss in a harmonic resonator is quantified by quality factor $Q$. Because independent dissipation channels add reciprocally, the total loss follows

$$\frac{1}{Q(p)} = \frac{1}{Q_{int}} + Ap^n \,. \tag{1}$$

Here, $Q_{\mathrm{int}}$ represents the pressure-independent quality factor ceiling of the device, while $Ap^n$ describes the pressure-dependent gas loss contribution. In the free-molecular regime, gas dissipation scales linearly with pressure and $n$=1. At higher pressures, the scaling can approach $p^{1/2}$ as viscous effects emerge, depending on the device geometry. An exponent fitted across multiple flow regimes is therefore interpreted as an effective value.

Whether gas damping is molecular or viscous is set by the Knudsen number $\mathrm{Kn}=\lambda/L$, where $\lambda=k_B T/(\sqrt{2}\pi d^2 p)$ is the mean free path of gas molecules with collision diameter $d$ at temperature $T$ and pressure $p$ with Boltzmann constant $k_B$. Free molecular, transitional, and continuum flow correspond to $\mathrm{Kn} \gg 1$, $\mathrm{Kn} \approx 1$, and $\mathrm{Kn} \ll 1$, respectively. For the out-of-plane X- Lamé mode studied here, the sidewall gap is not modulated by the resonant motion, so $L$ is the 150 μm width of the resonator body rather than the 10 μm sidewall gap. As the pressure rises, the equilibrium adsorbate coverage increases and loads the surface with mass, lowering $f_0$, until the finite population of adsorption sites saturate. When the pressure is stepped to a fixed value and held, the surface coverage relaxes toward a new equilibrium and need not equilibrate on the same timescale as gas damping. Consequently, $Q$ and $f_0$ probe distinct gas-resonator interactions and can exhibit markedly different temporal responses.

**Figure 2** presents the pressure-step response at 25.00 ± 0.04 °C as the pressure is stepped from 500 Torr to 0.01 Torr. Pressure is held constant at each level for 10 min as shown in **Figure 2**a, and the corresponding spectrum is recorded after 5 min using an automated acquisition sequence. The resonance-amplitude map in **Figure 2**b resolves two distinct modal branches, with the 64.21 MHz X-Lamé mode tracked throughout the experiment. The resonance frequency $f_0$ increases monotonically as the pressure decreases, producing a total shift of approximately 1.3 kHz that is consistent with desorption and reduced adsorbate mass loading, as shown in **Figure 2**c. The quality factor follows the reciprocal-loss model of Eq. (1) across more than four decades, with an intrinsic ceiling $Q_{\mathrm{int}} \approx 8.5 \times 10^4$ reached below approximately 1 Torr and a power-law gas term with an effective exponent $n \approx 1$. At 10 Torr, the fitted gas-loss contribution is approximately 1.5% of the pressure-independent loss and reaches 10% near 70 Torr. At 500 Torr, this contribution increases to approximately 70% of the pressure-independent loss, reducing $Q$ to approximately 60% of $Q_{\text{int}}$ (**Figure 2**d). Extrapolation of the fitted response places equal gas and pressure-independent losses near atmospheric pressure, where $Q=Q_{\text{int}}/2$. Because pressure-independent dissipation dominates most of the sweep, the measured $Q$ does not clearly resolve the distinct free-molecular, transitional, and continuum scaling regimes, even though the experiment spans all three. Weak gas coupling therefore allows $Q$ to remain near its intrinsic ceiling under moderate vacuum, whereas $f_0$ remains sensitive to pressure through surface mass loading.

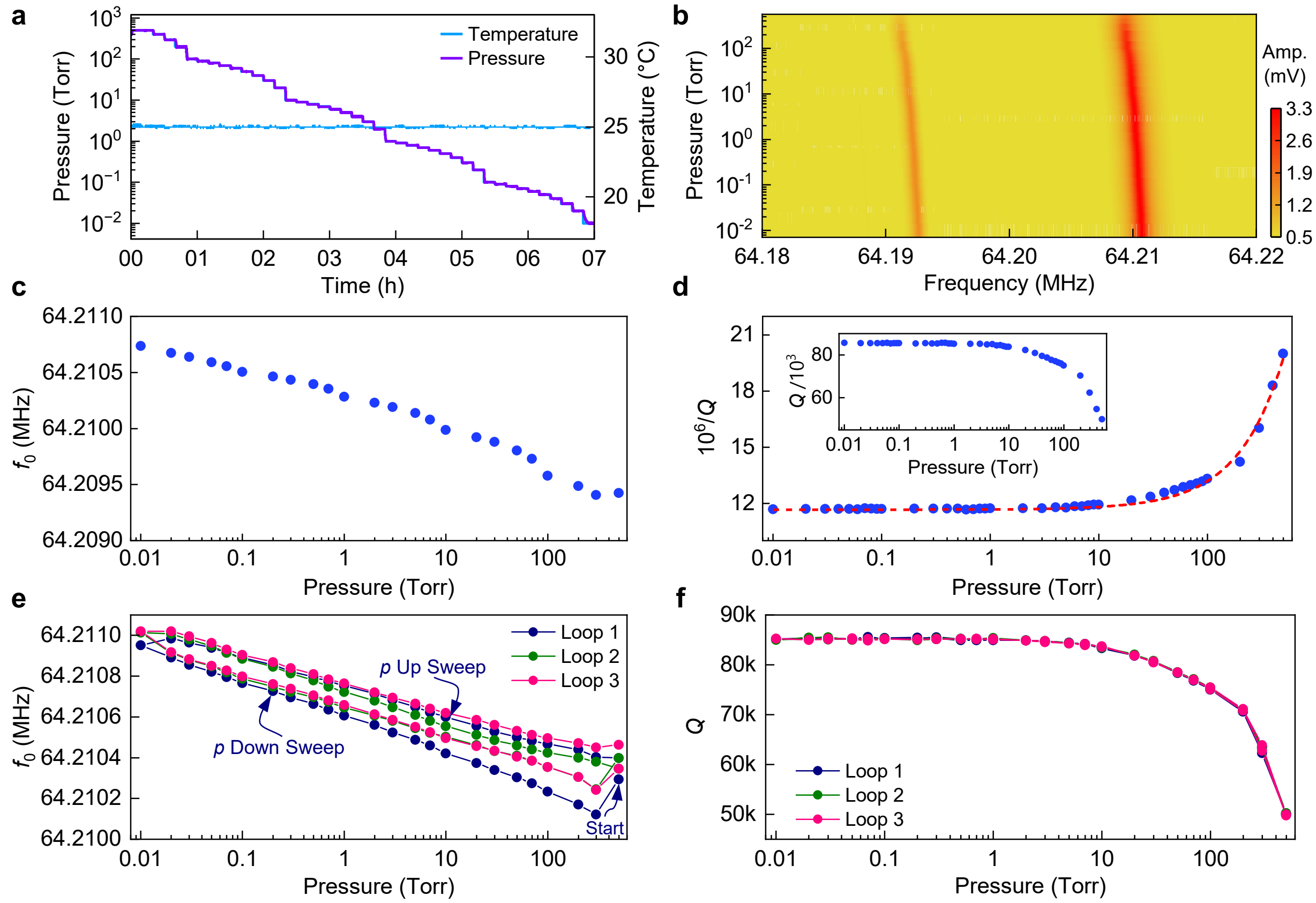

**Figure 2. Pressure-dependent $f_0$ and $Q$ at 25 °C.** (a) Applied pressure is stepped from 500 Torr to 0.01 Torr over time. The right axis shows that the measured temperature remains at 25 °C throughout the experiment. (b) Resonance amplitude color map resolving two nearby modes from distinct arms of the resonator. (c) $f_0$ rises monotonically as pressure falls, gaining approximately 1.3 kHz from 500 to 0.01 Torr. (d) Extracted $Q$ versus $p$ demonstrates that $Q$ remains near its intrinsic ceiling ($\approx 8.5 \times 10^4$) below ~1 Torr and decreases with increasing pressure, reaching approximately half this value near 500 Torr. (e) Three consecutive up-and-down pressure loops show that $f_0$ traces a repeatable hysteresis loop during each cycle, with the pressure down-sweep lying below the up-sweep and cycle-to-cycle variations limited to a few hundred Hz. (f) $Q$ traces overlap across all three loops with negligible hysteresis, indicating that pressure-dependent damping is reversible whereas $f_0$ retains a persistent dependence on pressure history.

Repeated pressure cycling probes the reversibility of these two channels (**Figure 2**e and **Figure 2**f). Across three consecutive pressure sweeps between 500 and 0.01 Torr, $f_0$ traces a hysteresis loop, with descending branch lying below the ascending branch, consistent with adsorption and desorption kinetics that do not equilibrate within the duration of each pressure step. The quality factor, in contrast, follows nearly identical trajectories during all three cycles, with negligible hysteresis, and returns to its intrinsic ceiling at low pressure, demonstrating that gas damping responds reversibly to the imposed pressure. This repeatability establishes $Q$ as a stable and reproducible measure of dissipation for this resonator and shows that the gas-damping channel carries no measurable lasting memory; the slow, history-dependent behavior is observed only in $f_0$.

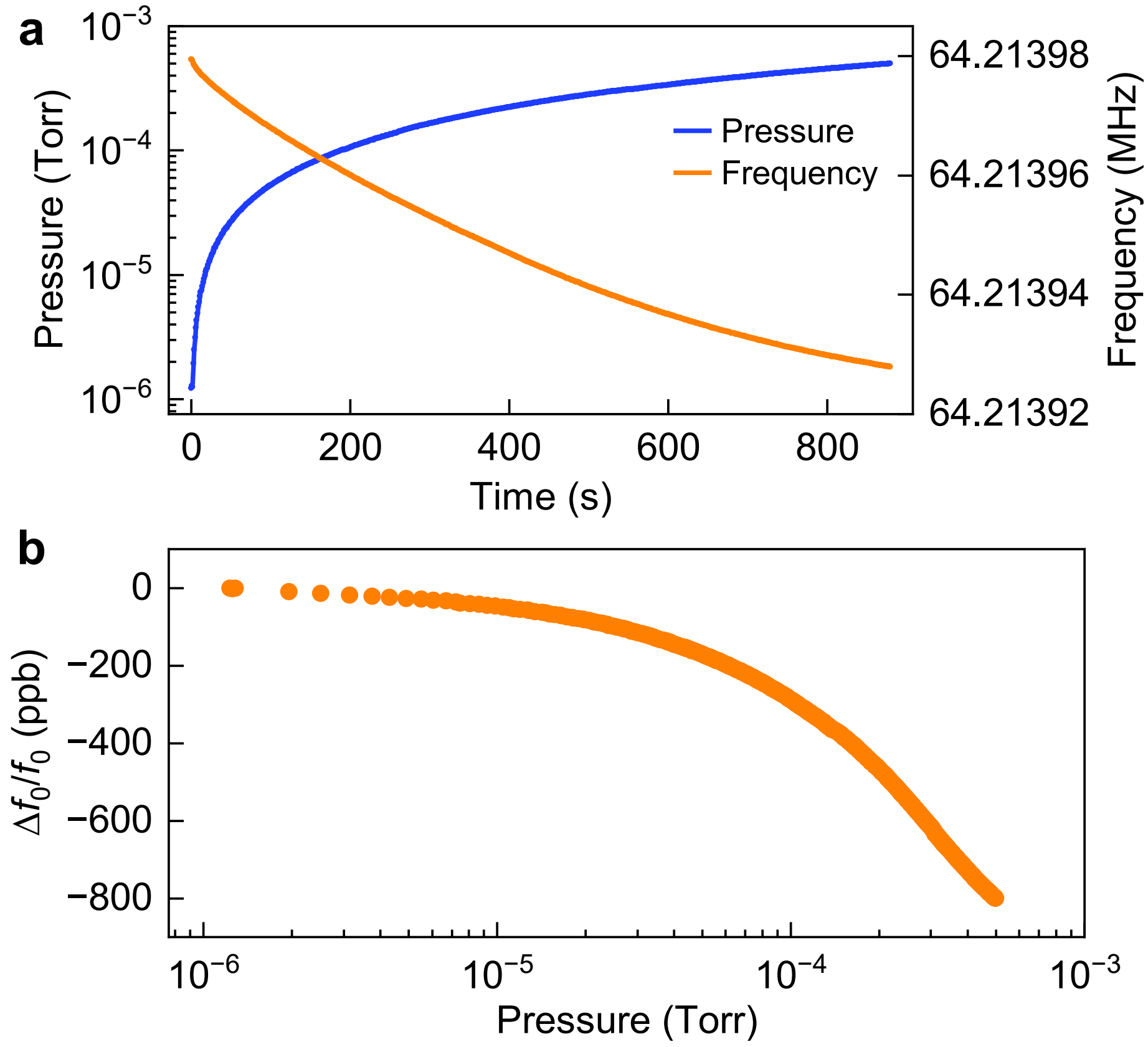

**Figure 3. High-vacuum frequency response measured at 25 °C.** (a) Chamber pressure and PLL-tracked resonance frequency $f_0$ during a pump-isolation transient. After 1 h of evacuation, the pump is isolated from the chamber, and the pressure rises from approximately $10^{-6}$ to $5\times10^{-4}$ Torr over 900 s while $f_0$ decreases monotonically. (b) Fractional frequency shift $\Delta f_0/f_0$ as a function of the simultaneously measured pressure. The fractional frequency shift remains nearly constant below approximately $10^{-5}$ Torr and reaches -800 ppb at $5\times10^{-4}$ Torr.

To investigate the frequency response in high vacuum, we evacuate the chamber with a turbo pump for 1 h and then isolate it from the pump while a phase-locked loop (PLL) continuously tracks $f_0$ (**Figure 3**). Following isolation, the chamber pressure rises from $10^{-6}$ to $5\times10^{-4}$ Torr over 900 s, while $f_0$ decreases monotonically (**Figure 3**a). The chamber temperature remains regulated at 25.00 °C throughout the transient, excluding a thermal origin for the shift. The decrease tracks the rising pressure and is consistent with rapid occupation of readily accessible adsorption sites. When the transient is plotted against the simultaneously measured pressure, the fractional frequency shift remains within approximately 50 ppb of its high-vacuum value below $10^{-5}$ Torr, after which the pressure dependence becomes more pronounced, reaching -800 ppb at $5\times10^{-4}$ Torr (**Figure 3**b). Because the pressure rise occurs much faster than the hour-scale equilibration established in the subsequent relaxation measurements, this trajectory represents a dynamic adsorption response rather than an equilibrium isotherm and likely underestimates the magnitude of the fully equilibrated frequency shift. The onset near $10^{-5}$ Torr places the pressure scale for a measurable frequency response at least five orders of magnitude below the approximately 1 Torr threshold at which $Q$ reaches its intrinsic limit. Thus, even when dissipation is no longer measurably sensitive to pressure, surface mass loading continues to impose a substantially stricter vacuum requirement on $f_0$.

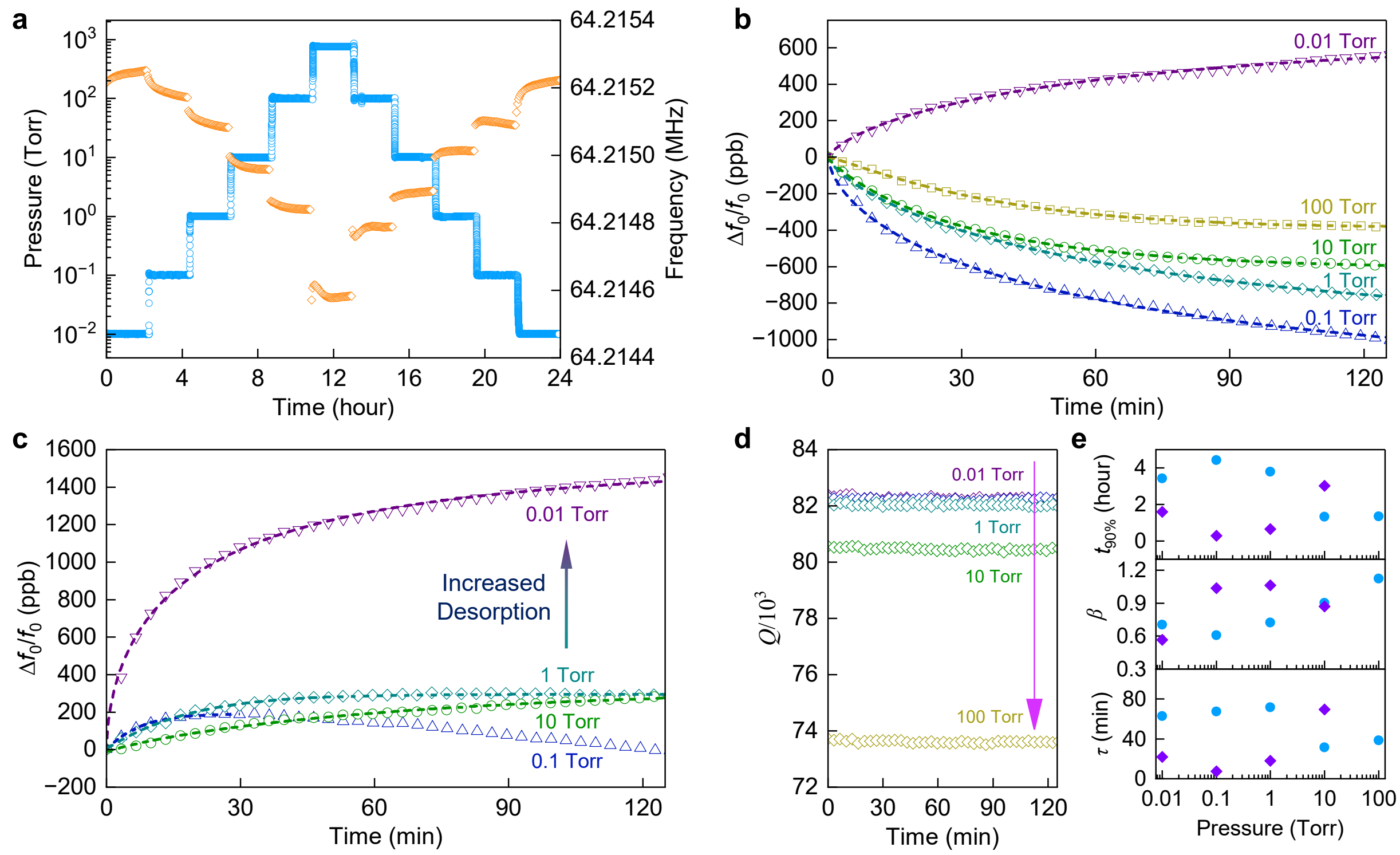

**Figure 4. Pressure-step relaxation kinetics at 68.8 C frequency turnover.** (a) Full pressure cycle and the simultaneously recorded $f_0$, with first-order temperature sensitivity suppressed. (b) On the pressure up-sweep, $\Delta f_0/f_0$ falls through adsorption-induced mass loading. Dashed curves show stretched-exponential fits, with larger relaxation amplitudes at lower pressures. The initial 0.01 Torr trace instead rises, reflecting desorption of adsorbates accumulated during the preceding atmospheric exposure. (c) On the pressure down-sweep, $\Delta f_0/f_0$ rises as adsorbate mass is released, with the relaxation amplitude increasing as pressure decreases and reaching approximately 1400 ppb at 0.01 Torr. (d) $Q$ varies by less than 0.3% during each hold, indicating that damping reaches steady state while $f_0$ continues to relax. (e) Extracted $t_{90}$, $\beta$, and $\tau$ versus pressure, where $t_{90} = \tau\,(\ln 10)^{1/\beta}$ is the time required for the fitted relaxation to reach 90% of its asymptotic frequency change. Blue-filled circles and violet filled squares denote the pressure up- and down-sweeps, respectively.

**Surface Kinetics Governing Frequency Relaxation**

To suppress first-order thermal contributions to the pressure-dependent frequency response, we perform the pressure-step measurements at the resonator's 68.8 °C turnover temperature (**Figure 4**a). Each pressure setpoint is maintained for 2 h 10 min while resonance spectra are acquired with the UHFLI. Raising the pressure drives net adsorption and $f_0$ falls through adsorption-induced mass loading (**Figure 4**b). Excluding the initial 0.01 Torr trace, the negative relaxation is largest at the lower pressure setpoints and approaches -1000 ppb at 0.1 Torr. The initial 0.01 Torr trace instead rises because it is the first hold following atmospheric exposure. Adsorbates retained on the surface desorb under vacuum, reducing the effective mass and raising $f_0$. On the down-sweep, every trace rises as the surface releases mass (**Figure 4**c), and the 0.01 Torr hold, reached after the high-pressure excursion, attains 1.4 ppm (≈ 90 Hz). The reversal in the direction of relaxation between the two sweeps is consistent with net adsorption during increasing pressure and net desorption during decreasing pressure. A single pressure step therefore produces a fractional frequency shift as large as 1.4 ppm, comparable to a year of aging in commercial quartz and MEMS oscillator.[20] After the chamber pressure reaches the setpoint, $Q$ varies by less than 0.3% throughout the hold (**Figure 4**d), whereas $f_0$ continues to evolve by tens of hertz over the same interval. Damping therefore reaches steady state on the pressure control timescale, while the frequency retains a slow dependence on the evolving surface state.

The shape of each relaxation carries more information than its amplitude. A single exponential does not reproduce the progressively slowing tails, so each transient is fitted to the stretched-exponential form

$$f_0(t) = f_\infty + Be^{-\left(\frac{t}{\tau}\right)^{\beta}}, \tag{2}$$

where $f_\infty$ is the asymptotic resonance frequency, $B$ is the signed relaxation amplitude, $\tau$ is the characteristic time, and $0 < \beta \leq 1$ is the stretching exponent. When $\beta = 1$, Eq. (2) reduces to relaxation governed by a single characteristic time, whereas $\beta < 1$ corresponds to a distribution of relaxation rates. The recovered exponents lie consistently below unity for both adsorption and desorption at every pressure (**Figure 4**e), consistent with an energetically heterogeneous surface supporting a broad range of kinetic rates. Over a finite observation interval, such a broad relaxation spectrum can also appear approximately logarithmic, providing a physical connection to the empirical aging laws used for precision oscillators.[20]

The finite 2 h 10 min observation window does not fully resolve the slowest components and can bias $\tau$ downward and $\beta$ upward. The values in **Figure 4**e are therefore apparent parameters with $\beta \approx 0.5$ to 0.8, whereas the 10 h records of **Figure 4** yield $\beta \approx 0.4$. Expressed as the time to reach 90% of the asymptotic frequency change, $t_{90} = \tau \cdot (\ln 10)^{1/\beta}$, $t_{90}$ exceeds the 10 min dwell time of the **Figure 2** sweeps at every pressure, providing a quantitative basis for the observed hysteresis in $f_0$.

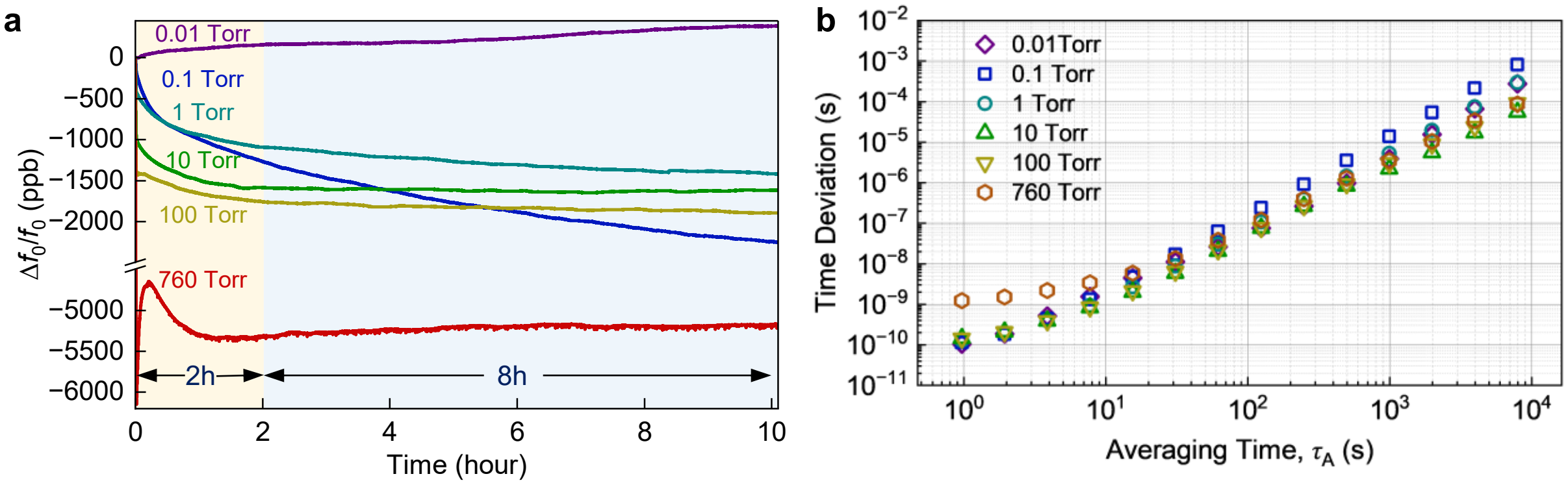

**Figure 5. Closed-loop frequency aging and time-deviation behavior across the pressure ladder at the 68.8 °C turnover.** (a) Fractional frequency shift recorded by a phase-locked loop for 10 h at each pressure of the ascending ladder from 0.01 to 760 Torr, with each trace zeroed to its own value at the start of the hold. Each upward pressure step drives a decelerating adsorption relaxation, deepest and slowest at 0.1 Torr, where $f_0$ falls 1900 ppb and is still relaxing at 10 h. The first hold at 0.01 Torr instead rises by 250 ppb as adsorbate loaded during air exposure desorb, and the 760 Torr hold falls rapidly to -6200ppb, then overshoots by 700 ppb before partially recovering, consistent with reorganization of the adsorbed layer. (b) Time deviation computed from the final 8 h of each hold, excluding the post-step transient and without detrending. All subatmospheric holds collapse onto a common floor of $\sigma_x \approx 1.0$ to $1.4 \times 10^{-10}$ s at $\tau_A = 1$ s and a shared flicker branch up to $10^3$ s, insensitive to pressure across five decades. The 760 Torr floor sits tenfold higher, consistent with the halved $Q$ and reduced resonance amplitude, and the upturn beyond $10^3$ s carries the deterministic aging of (a) rather than added stochastic noise.

**Pressure-Dependent Frequency Aging and Stability**

**Figure 5** examines how the surface kinetics affect the long-term frequency stability of the resonator under closed-loop tracking. The PLL continuously tracks $f_0$ for 10 h at each pressure in an ascending sequence from 0.01 to 760 Torr at 68.8 °C, with each trace referenced to its value at the start of the hold (**Figure 5**a). During the first hold at 0.01 Torr, $f_0$ rises by 250 ppb, consistent with net desorption following atmospheric exposure, whereas every subsequent pressure increase initially causes $f_0$ to fall. At 0.1 Torr, $f_0$ shows the largest response, falling by 1900 ppb, the largest single-step shift observed in this work, and remaining measurably nonstationary at the end of the 10 h measurement. The shifts at intermediate pressures are smaller and approach steady values more rapidly. At 760 Torr, the first 2 min of the record show a rapid decrease of approximately 6200 ppb, followed by a rebound and a second, slower decrease. These measurements establish that $f_0$ can remain nonstationary for hours after the chamber pressure has stabilized, without assigning the relaxation to a specific microscopic pathway. A thermal origin is excluded by the operating point itself, since $\mathrm{d}f_0/\mathrm{d}T$ vanishes at the 68.8 °C turnover.

**Figure 5**b presents the time deviation (TDEV), $\sigma_x(\tau_A)$, computed from the final 8 h of each 10 h record, excluding the post-step transient; no detrending is applied, so the TDEV retains the residual deterministic relaxation and characterizes the combined resonator-readout-PLL response rather than the intrinsic noise of the mechanical resonator alone (Methods).

At an averaging time of $\tau_A \approx 1$ s, the subatmospheric measurements from 0.01 to 100 Torr cluster between $1.0 \times 10^{-10}$ and $1.4 \times 10^{-10}$ s, indicating weak pressure dependence in the measured short-term stability. The 760 Torr measurement exhibits a TDEV of approximately $1.2 \times 10^{-9}$ at $\tau_A \approx 1$ s, nearly one order of magnitude above the subatmospheric measurements. Characterization at 68.8 °C and a fixed drive of -20 dBm gives $Q \approx 40000$ at 760 Torr and $Q \approx 85000$ at 0.01 Torr. The resulting broader and lower-amplitude resonance at 760 Torr provides poorer frequency discrimination to the PLL and is consistent with the elevated short-term TDEV.

Beyond approximately $10^3$ s, the pressure associated with the largest TDEV shifts from 760 Torr to the low-pressure measurements. The 0.01, 0.1, and 1 Torr frequency traces continue to evolve, as shown in **Figure 5**a, and exhibit the largest TDEV at longer $\tau_A$. Because the deterministic trend is retained in the TDEV calculation, this increase should not be assigned to a stationary noise process. **Figure 5**b therefore reveals a crossover between atmospheric gas-damping degradation of short-term closed-loop tracking and surface-related relaxation that limits long-term frequency stability at lower pressures.

## 3. Conclusions

In summary, we have demonstrated that dissipation equilibrium and frequency equilibrium represent distinct states in a mechanical resonator exposed to a fixed gaseous environment. Gas-mediated loss stabilizes rapidly after a pressure change, whereas the adsorbate population continues to evolve, producing deterministic frequency aging after the cavity pressure has become stationary. The absence of a corresponding pressure dependence in random frequency fluctuations further identifies this aging as a slow surface process rather than pressure-induced frequency noise. Thus, $Q$ remains a valid measure of damping and can indicate cavity pressure within its pressure sensitive regime, but it cannot establish whether $f_0$ has equilibrated. Vacuum-package specifications for long-term frequency stability should therefore include the allowable evolution of $f_0$ over the relevant operating interval. Hence, as resonators are miniaturized, controlling surface chemistry and adsorbate populations will become as important as reducing cavity pressure.

## 4. Methods

### Resonator Fabrication Process

The resonator is fabricated on a silicon-on-insulator wafer comprising a 50 μm degenerately phosphorus-doped single-crystal silicon device layer, a 3 μm buried oxide, and a 500 μm handle layer. The device layer forms both the mechanical resonator body and the electrically conductive bottom electrode. A 600 nm c-axis-oriented AlScN film is deposited onto the device layer by reactive magnetron sputtering. A 50 nm molybdenum layer is subsequently sputtered and patterned by $SF_6$-based reactive-ion etching to define separate drive and sense electrodes. $Cl_2$-based reactive-ion etching then locally removes the AlScN film, exposing the doped silicon for electrical contact to the bottom electrode. Metal routing and contact pads are formed by lifting off a 10 nm chromium adhesion layer beneath 300 nm of platinum, thereby establishing low-resistance contacts to the device layer. A Bosch deep reactive-ion etch through the full device layer to the buried oxide simultaneously defines the resonator body, the 10 μm isolation gap, and the phononic-crystal tethers. The etch conditions are controlled to minimize sidewall roughness and overetching at the $Si/SiO_2$ interface, and residual polymers are subsequently removed by plasma ashing. The handle layer is etched from the backside to the buried oxide, after which selective removal of the exposed oxide releases the resonator. The buried oxide acts as an etch stop during silicon etching, preserving a smooth bottom surface.

**Time-Deviation Analysis of PLL-Tracked Frequency**

At each pressure, the UHFLI PLL continuously tracks $f_0$ for 10 h. The first 2 h following each pressure transition are excluded to reduce the contribution of the initial frequency excursion, and the final 8 h are used for time-deviation analysis. Each frequency sample $f_i$ is converted to fractional-frequency deviation according to

$$y_i = \frac{f_i - \bar{f}}{\bar{f}} \tag{3}$$

where $\bar{f}$ is the mean frequency over the selected 8 h interval. The effective sampling interval $\tau_0$ is determined from the median separation between consecutive timestamps. TDEV is calculated at octave-spaced averaging times $\tau_A = m\tau_0$ using the tdev() function in AllanTools with the input specified as fractional-frequency data. TDEV is related to the modified Allan deviation by [32]

$$\sigma_x(\tau_A) = \frac{\tau_A}{\sqrt{3}} Mo\,d\,\sigma_y(\tau_A)\,, \tag{4}$$

where $Mod\,\sigma_y(\tau_A)$is the modified Allan deviation and $\sigma_x(\tau_A)$ is expressed in seconds. No linear or nonlinear detrending is applied to the selected frequency traces. The calculated TDEV therefore includes both stochastic fluctuations and the residual deterministic relaxation present during the final 8 h. It characterizes the combined response of the resonator, UHFLI readout, and PLL and is not interpreted as the intrinsic resonator noise alone.

## Acknowledgements

This work was financially supported by the Defense Advanced Research Projects Agency (DARPA) H6 (HR00112390018) and NIMBUS (HR00112590104) programs. The authors would like to thank the University of Michigan Lurie Nanofabrication Facility (LNF) cleanroom staff for fabrication support. The views, opinions and/or findings expressed are those of the authors and should not be interpreted as representing the official views or policies of DARPA or the U.S. Government.

**Notes:** The authors declare no competing financial interest.